\documentclass[useAMS,usenatbib]{mn2e}
\def\b{\begin{equation}}
\def\e{\end{equation}}
\def\l{\left}
\def\r{\right}
\title[  Eternally Collapsing Objects ]
{ Radiation Pressure Supported Stars in Einstein Gravity: Eternally
Collapsing Objects}
\author[A. Mitra ]{A. Mitra$^1$\thanks{On leave from  BARC, Mumbai-400085
India,
Abhas.Mitra@mpi-hd.mpg.de}\\
$^1$ Max Planck Institut fur Kernphysik, Saupfercheckweg 1, D-69117
Heidelberg, Germany\\}

\begin{document}

\date{Accepted 2006 March 16. Received  2006 March 15; in original form 2006 January 13 }

\pagerange{\pageref{492}--\pageref{496}} \pubyear{2006}

\maketitle

\label{firstpage}

\begin{abstract}
Even when we consider Newtonian stars, i.e., stars with surface
gravitational redshift, $z \ll 1$, it is well known that,
theoretically, it is possible to have stars, supported against
self-gravity, almost entirely by radiation pressure. However, such
Newtonian stars must necessarily be supermassive\citep{b1, b2, b3}.
 We point out that this requirement
for excessive large $M$ in Newtonian case, is a  consequence of the
occurrence of low $z\ll 1$. But if we remove such restrictions, and
allow for possible occurrence  of highly general relativistic
regime, $z \gg 1$, we show that, it is possible to have radiation
pressure supported stars at arbitrary value of $M$. Since radiation
pressure supported stars necessarily radiate at the Eddington limit,
in Einstein gravity, they are never in strict hydrodynamical
equilibrium. Further, it is believed that sufficiently massive or
dense objects undergo continued gravitational collapse to the Black
Hole stage characterized by $z =\infty$. Thus, late stages of Black
Hole formation, by definition, will have, $z \gg 1$, and hence would
be examples of quasi-stable general relativistic radiation pressure
supported stars \citep{b4}. It is shown that the observed duration
of such Eddington limited radiation pressure dominates states is $t
\approx 5\times 10^8 (1+z)$ yr. Thus,  $t \to \infty$ as Black Hole
formation ($z\to \infty$) would take place. Consequently, such
radiation pressure dominated extreme general relativistic stars
become Eternally Collapsing Objects (ECOs) and and  the BH state is
preceded by such an  ECO phase. This result is also supported by our
previous finding that trapped surfaces are not formed in
gravitational collapse\citep{b5} and the value of the integration
constant in the vacuum Schwarzschild solution is zero\citep{b6}.
Hence the supposed observed BHs are actually ECOs.
\end{abstract}

\section{Introduction}
Stars may be defined as objects with an intrinsic self-luminosity
which is generally sustained by the grip of self-gravity. Most of
the stars that we know of are primarily supported by gas pressure
$p_g$ rather than by radiation pressure $p_r$. The usually
insignificant role of $p_r$ is indicated by the parameter
\citep{b7}: \b \beta = {p_g\over p}
 \e
 where $p = p_g
+ p_r$ is the total pressure. For most of the known luminous
(Newtonian) stars $\beta \approx 1$. The actual role of $p_r$ is
seen more clearly from the adjoint parameter: \b 1 - \beta =
{p_r\over p} \e The relative role of $p_r$ may even be  more
directly expressed through the parameter
\b
 x \equiv {p_r\over p_g}
={1 -\beta \over \beta} \e For real astrophysical objects, $x$ would
be highest at the centre and and in general decrease significantly
towards the surface.  The central value of $x^*=0.006$ for the Sun
and the same for a $ 4 M_\odot$ mainsequence star is 0.012 (as
kindly pointed out by the anonymous referee) ($\odot$ denotes solar
values).
 It may be borne in mind that
 the mean value of $(1- \beta) \ll (1- \beta^*)$. It is known, however, that the value of $(1 -\beta^*)$
would rise significantly if the value of $M$ would be very high
though we hardly know of any single Newtonian star with $M > 100
M_\odot$. In fact it may  be pointed out that in any given stellar
model, it is the value of $(1- \beta^*)$ which determines the mass
of the star \citep{b7}. Suppose one makes stellar models with a
fixed composition ($\mu$). Then under the assumption of fixed
chemical composition, a stellar mass of $M$ will get defined from
the equation \citep{b7} (p. 82):
 \b
 {1- \beta^*\over {\beta^*}^4} = {{1- \beta_\odot} \over
\beta_{\odot}^4} \l({M\over M_\odot}\r)^2
 \e
  so that $M$ slowly
increases with decreasing $\beta^*$ (incresing $x$). Note $M \to 0$
if $\beta^*\to 1$ or $x^* \to 0$ in this non-quantum  case, and to
have a finite mass luminous star, one must have a finite value of
$x$. Even in the case of (partially) degenerate objects, a low value
of $\beta$, or a higher value of $x$ would
 raise their  masses. The usual Chandrasekhar mass, $M_c$ corresponds
  to a perfectly degenerate
object having ultimate relativistic degeneracy where momentum of the
pressure giving particles $P \to \infty$ and further the object is
absolutely cold, $T =0$. Moreover, it may be reminded that, {\em the
radius of Chandrasekhar's critical white dwarf is zero}, $R_c =0$!
This shows that, only an infinitely dense singular configuration
could be perfectly degenerate and have $T=0$, and, on the other
hand,  all real finite configurations, in the strict sense, must be
either partially degenerate or at a finite $T$. When the compact
object has a finite temperature despite having assumed degeneracy,
the mass {\em upper limit gets modified} as \citep{b7} (see p. 437):
 \b
  {\cal M}_c = M_c \beta^{-3/2}
  \e
  In
order to keep the object  almost ``completely degenerate'',
Chandrasekhar imposed some artificial restrictions on the value of
$\beta$ which resulted in \b
 {\cal M}_c = 1.156 M_c
 \e
 But the fact
remains that perfect degeneracy, in a strict sense, corresponds to
$\beta =1.0$ or $x=0$. It may be noted that the {\em radius of this
almost completely degenerate white dwarf too is zero}, ${\cal R}_c
=0$ \citep{b7} (see p. 441, eq. 140). However, theoretically, the
possibility of having arbitrary finite $x$ cannot be ruled out. As
value of $x$ would rise, the system would be more and more partially
degenerate and soon, it would be more meaningful to call it
``non-degenerate''. In such a case, as $\beta \to 0$, one may have
${\cal M}_c \to \infty$, i.e., {\em the very notion of an ``upper
mass limit'' would vanish}. Actually, the applicability of Eq.(5)
would cease once degree of degeneracy would significantly reduce.

As indicated by Eq.(4), in principle, an initially nondegenerate
sequence of stars can always be turned into radiation pressure
supported stars (RPSSs) by considering appropriate higher value of
$M$ and vice-versa. Thus, irrespective of whether one follows the
standard non-degenerate route or the degenerate route, one may, in
principle,  end up with non-degenerate radiation pressure supported
stars. Now we focus attention on the question of mass range of such
radiation pressure supported stars (RPSS) in the context of both
Newtonian  and Einstein gravity.
\section{ General Formalism}
In reality, even if one would consider the stellar material to be pure hydrogen,
at sufficiently high temperature, one would have pressure contribution from both
electrons and protons. Also there will always be rest mass contribution by the electrons. But in the following, since we are interested only in seeing some new qualitative
result, we will consider an idealized fluid which is assumed to
always remain  monoatomic.
The total proper mass energy density of this fluid is
\b
\rho = \rho_g + (\gamma -1)^{-1} p_g + \rho_r
\e
where
\b
\rho_g = m_p n c^2
\e
is the (baryonic) rest mass energy density, $n$ is the proper baryonic number density,
 $m_p$ is proton rest mass,
$c$ is the speed of light, and $\gamma$ is the  ratio of specific
heats. Also \b p_g = n k T \e is the gas pressure, where $k$ is
Boltzmann constant and \b \rho_r =  a T^4 \e is the radiation energy
density. Here $a=  7.56 \times 10^{-15}$ erg cm$^{-3}$ $K^{-4}$ is
the radiation constant.
 As is known,
\b
p_r = {1\over 3} a T^4
\e
From Eqs. (9) and (11), we may also
find that
 \b
  x = {p_r\over p_g} = {a T^3 \over 3 n k}
  \e
   Further,
instead of $\beta$, it will be more convenient  define a quantity
$\beta_w = x^{-1} = p_g/p_r$. One may also define a related
parameter
 \b
y = {\rho_r\over \rho_g} = {a T^4 \over m_p n c^2}
\e
 which is the
ratio of the radiation energy density and the rest mass energy
density. In Newtonian gravity, one has  $\rho_r \ll \rho_g$, i.e.,
$y \ll 1$, and, consequently, $\rho \approx \rho_g$. In idealized
models, fluids may be characterized by a polytropic equation of
state (EOS): \b p = K \rho^{1 + 1/m} \e where $K$ must be uniform
within the fluid.
  If there would exist
a {\em strictly static} self-gravitating configuration, where, $K$ and $m$ would be
 same everywhere, then one would have a polytrope of degree ``$m$''.
 For the construction of stellar models, one looks for polytropes which have a finite boundary
 radius $R$. Mathematically, this would mean that, $\rho = p =0$ for $r > R$.
 Both in Newtonian and GR case, in a strict sense, {\em this vital boundary condition
 can be satisfied only if the object is absolutely cold} $T=0$ so that
 $p_r =\rho_r=0$ for $r > R$ and because one may indeed make $p_g =\rho_g=0$ at the boundary. In Newtonian gravity, since,
 $\rho_r = 0$, and $\rho =\rho_g$, this boundary condition can be satisfied much more accurately if not absolutely.
  In Newtonian gravity,  one can consider
 the results obtained by static polytropic models to be reasonably accurate. Now we shall
 attempt to qualitatively understand why Newtonian RPSSs must be of extremely high mass.
 \subsection{ RPSSs in Newtonian Gravity}
 For the best known low mass Newtonian star, namely the Sun, as we know,
 the {\em central value} of
 $\beta_* \approx \beta_w \approx 0.003$. Since $T$ drops sharply as one moves
 away from the centre, the mean value of $\beta$ must be extremely low
 in comparison to the above value. One may estimate the mean value of $x$
 for low mass Newtonian stars in the following way.
 By using Newtonian virial theorem for Newtonian stars, one can estimate the
 internal energy of the star as\citep{b7}
 \b
 U = {1\over (5\gamma -6)} {G M^2 \over R}
 \e
 where $\gamma$ is the effective ratio of specific heats, $G$ is the gravitational
 constant and $R$ is the radius of the star. As long as
 radiation pressure is really small, one can have $\gamma \approx 5/3$.
 Further, if the mean temperature of the assumed monoatomic gas is $T_m$,
 one has
 \b
 U = {3\over 2} {M\over m_p} ~k T_m
 \e
 Then by combining Eqs. (15) and (16), it follows that
 \b
 k T_m = {2\over 7} {G M m_p\over R }
 \e
 In terms of  the surface redshift or compactness $z = GM /R c^2$,
  we can rewrite the mean temperature as
 \b
 k T_m = {2 \over 7} ~z~ m_p c^2
 \e
 We may recall a that for Sun,
 $z \approx 2\times 10^{-6}$.
 This shows that the temperature of Newtonian stars are necessarily low
 {\em because they are hardly compact}, i.e., $z \ll 1$.  Alternatively, since, $k T_m$
 is low in comparison to nucleon rest mass energy  $m_p c^2$, radiation pressure is low. In other words, radiation pressure is
 relatively low  (for low $M$ stars) because compactness is so low.
 One can ask then, what is the appropriate mean value of $x$ for  low mass
 Newtonian stars. To see this we invoke a relationship which shows that
 for $z \ll 1$, one has\citep{b4}
 \begin{equation}
y  \sim \alpha ~{G M \over R c^2} \sim \alpha ~z
\end{equation}
where
 \b
  \alpha = {L\over L_{ed}}
\e
 is the luminosity of the star
in terms of its Eddington luminosity $L_{ed}$.
 By combining Eqs. (12), (13) and (19), we find that
 \begin{equation}
x  \approx \alpha ~z~ { m_p c^2\over 3 k T}
\end{equation}
Since we are interested only in the gross mean properties rather
than precise central quantities, in the foregoing equation, we may
consider $T= T_m$ because virial temperature is a reasonably good
measure of the gross mean temperature. Then we may use Eq.(18) in
Eq.(21) to
 find that
 \b
 x \approx {7\alpha\over 6} \approx \alpha
 \e
 For Sun, $L_{ed}$, at the surface is $1.3 \times \sim 10^{38}$ erg/s so that
 $\alpha \sim 3\times 10^{-5}$. Therefore the mean value of $x$ for Sun is also
 $\sim 3 \times 10^{-5}$. Again recall that,  $z \approx 2\times 10^{-6}$ for the Sun. Note the cosmic coincidence that, for Sun, the three apparently uncorrelated
 dimensionless quantities, namely, $x$, $\alpha$ and $z$  have similar values within a factor of $10$!
 In order to become a RPSS, the value of mean
 $x$ has to rise from such an extreme low value to attain a stage of $x \gg 1$.
Since for  low mass stars, $L \sim M^{5.5}/R^{0.5}$ while $L_{ed}
\sim M$, we will have,
 $\alpha = L/L_{ed} \sim M^{4.5}/R^{0.5}$. For low mass main-sequence stars, crudely,
 $M \propto R$, so that, $\alpha \sim M^4$. Thus if one would move to higher mass
 stars, initially, mean $x$ would increase very rapidly.
 However, as $x$ increases, Eq.(15) would need significant revision  as the star
 would move away from being a gas pressure dominated $\gamma \approx 5/3$ polytrope
 to a radiation pressure dominated $\gamma \approx 4/3$ polytrope. In such a case, the pattern of
 variation of $x$ with $M$ would change drastically.
 It appears,  that, in the regime of $p_r \gg p_g$, the supposed Newtonian static RPSSs  may be reasonably described
 by a Newtonian polytrope with $m \approx 3$\citep{b3}.  Then,  the mass of the
 star is\citep{b3}
 \b
 M = 18 M_{\odot} {\beta_w}^{-2} ~ \mu^{2} = 18 M_{\odot}
 ~x^2~\mu^2
 \e
 where $\mu$ is the mean molecular weight. For a fully
 ionized hydrogen plasma $\mu =2$
 It is interesting to invert Eq.(23) as
 \b
 x \approx 0.2 \left({M\over M_{\odot}}\right)^{1/2}~\mu^{-1}
\e
 which shows that in the large $M$ range $x$ increases relatively more rapidly as $M^{1/2}$ for the Newtonian stars. And eventually, still, $M$ has to be very high
 to ensure $x \gg 1$.
 To have a value of $x \sim 10$ (atleast) one must have, $M \sim 1800 M_\odot ~\mu^2$ which
is already very high. Note that, since, in principle, $x$ can be
arbitrarily high, again, there is {\em no upper limit on the mass of
self-gravitating configurations once we allow existence of
sufficiently high radiation pressure}. We may also see how the
``compactness'' of the Newtonian star would change in this very high
$M$ range. To appreciate this point further, we consider specific
model of Newtonian RPSSs\citep{b3} (see Eq. [11.5.9]): \b
 R \propto
\beta_w^{-2/3} \propto x^{2/3}
\e
Using  Eqs. (23) and (25), we find
then, that, in Newtonian case
\b
 z \propto M/R \propto x^{4/3}
 \e
Further, using Eq.(24) into the foregoing equation, we see that
 \b
  z \propto (M^{1/2})^{4/3} \propto
M^{2/3}
 \e
 also increases quite rapidly with mass. Thus compared to
the low mass Newtonian stars, in the high mass range, {\em on a
relative scale}, one can have quite high values of $z$. However,
this does not mean that for Newtonian stars one can ever have $z \gg
1$. This point has been emphasized by Weinberg:

$\bullet$ The entire derivation leading to the concept of any
Newtonian star necessarily assumes $y \equiv \rho_r/\rho_g \ll 1$.
And this assumption must be ensured for (Newtonian) supermassive
stars as well. As shown by Weinberg, in such a case, one must always
have\citep{b3}
\b
z = {GM \over R c^2} \ll 0.39
 \e
  In summary, the
{\em fundamental reason that Newtonian RPSSs stars are extremely
massive is that, by definition},  $z \ll 1$, and, in particular,  $z
\ll 0.39$. It may be also pointed out that since in general $L
\propto M^d$, where, $d >1$ (for low mass stars, $d \sim 5$), and
$L_{ed} \propto M$, $\alpha$ in general increases with $M$. Then
Eq.(27) would show that, atleast, in the Newtonian regime, in
general, $\alpha$ increases with $z$. However, the maximum value of
$\alpha =1$, and all the RPSSs, Newtonian or Relativistic,  have
this maximal value of $\alpha$ because a larger $\alpha$ would
disrupt the star due to excessive radiation pressure.
\section{Radiation Pressure Supported Stars  In Einstein Gravity}
The general definition of ``compactness'' in GR may be given in
terms of the surface redshift
\begin{equation}
 z= (1- 2 GM/R c^2)^{-1/2} -1
 \end{equation}
 One then easily finds
 that when $GM/R c^2 \ll 1$, (i.e., in the truly Newtonian regime), one has
 \begin{equation}
 z \approx GM/R c^2
 \end{equation}
It also follows that, the Event Horizon of a Black Hole (BH),
defined by $R = R_s= 2GM/c^2$ corresponds to $z=\infty$. And this
explains why gravitation is extremely strong for BHs. As is well
known,  very massive objects  undergo continued collapse to become
Black Holes. Thus, by definition, very massive objects, during their
continued collapse, must pass through stages having $z
> any~finite~number$. Although, it is not necessary, we may
nonetheless mention that the exterior spacetime of any contracting
self-gravitating object is represented by {\em radiating} Vaidya
metric\citep{b8} which allows the possibility that $z \to \infty$.
Further, it has been shown that, whenever, self-luminous objects
have $z\gg 1$\citep{b4}, one will have
\begin{equation}
y  \sim \alpha~ z/2
\end{equation}
Then by combining Eqs. (12) and (13),  Eq.(22) gets modified, in the
extreme GR case, as
\begin{equation}
x  \sim \alpha ~(z/2)~ { m_p c^2\over 3 k T}
\end{equation}
In all physically realistic cases of self-luminous objects, $\alpha$
is always finite. In the Newtonian case of $z\ll 1$, we found that,
$\alpha$ increases rapidly with $M$ and hence with $z$. Since
increase of $z$ implies stronger self-gravity, we may say that as if
{\em stronger gravity churns out more radiation from  self-luminous
self-gravitating objects}. And as $z\to \infty$, the entire object
becomes a ball of radiation/pure energy ($\rho_r/\rho_g \to
\infty$). Note that  this happens before the formation of any Event
Horizon(EH) and which indicates that the EH is actually synonymous
with the central singularity and the integration constant of the
vacuum Schwarzschild solution has the unique value of zero\citep{b5,
b6}. Actually, one can have continued gravitation collapse for
arbitrary (small) $M$ too provided $\rho$ would be suitably high. In
contrast to the Newtonian case, where a (relatively) higher $z$
demands, higher $M$ (Eq.[27]), in GR, the very fact that there could
be continued gravitational collapse at any mass scale means that one
can have high $z \gg 1$ {\em for arbitrary mass} and hence, one may
have $\alpha \to 1$, at suitably high value of $z$, at arbitrary
mass scale. For instance, for the Sun, $\alpha \approx 3\times \sim
10^{-5}$ and $z \approx 2 \times 10^{-6}$. But if we consider a
collapsing hot proto-neutron star in its final stage when it is
giving birth to a hot neutron star, $\alpha \approx 10^{-3}$ as $z
\sim 10^{-1}$\citep{b4}. Note the rise in the value of $\alpha$ with
$z$ even at a low mass scale of $M \sim 1 M_\odot$ as we are about
to enter the regime of Einstein gravity.

The gravitational mass energy of the star is defined as
\b
M c^2=
\int_0^R 4 \pi r^2 ~\rho ~ dr
 \e
  The mean value of mass energy
density is thus \b
 \rho = {3M c^2\over 4 \pi R^3}
 \e
 From Eq.(29),
note that in the $z \gg 1$ range,  the radius of the contracting
body would be hovering around its instantaneous Schwarzschild
radius, i.e.,  $R \approx R_s = 2 G M/c^2$. Using this  fact in the
foregoing equation, we have
\b
 \rho = { 3 c^8\over 32 \pi G^3 ~M^2}
\e
 Note that Eq.(31) implies that, for $z \gg 1$, $\rho_r \gg
\rho_g$ so that total $\rho = \rho_r + \rho_g \approx \rho_r$.
Further recalling that $\rho_r = a T^4$, from Eq.(35), we obtain
 \b
T = \l({ 3 c^8\over 32 \pi ~a~G^3~ M^2}\r)^{1/4} \e Numerically, one
finds, that, for such a state,  the mean temperature of the body in
this phase is :
 \b
 T \sim 600 ~\l({M \over M_\odot}\r)^{-1/2} ~MeV
\e
By substituting Eq.(37) in Eq.(32), we obtain
\b x \sim \alpha ~
z ~\l({M \over 15 M_\odot}\r)^{1/2}
 \e
 Note that, even in the extreme GR case, $x$ increases with $M$, and, in particular
 the behaviour $x \propto M^{1/2}$ is just what we found for Newtonian RPSSs. But unlike the Newtonian case, in the GR case, one can have $z \gg 1$. Eq.(38) shows that, with $\alpha >0$, and, $z \to \infty $,
one would be able to satisfy the condition $x \gg 1$ at arbitrary
mass scale for appropriate high value of $z \gg 1$. As mentioned
earlier, occurrence of $x \gg 1$ means $\alpha =1$ (rather than
$\alpha \gg 1$, which would disrupt the star). For isolated bodies,
radiation pressure and energy density are directly associated with
outward radiation/heat flux. This outward radiation flow has a
repulsive action on the plasma. As radiation pressure tends to grow
unabated with unabated increase of $z$, repulsive effect on the
plasma  must be able to  counterbalance the pull of gravity at some
stage. And this stage corresponds to attainment of $L\approx L_{ed}$
or $\alpha \approx 1$. In fact, domination of radiation pressure is
a much less demanding phenomenon than domination of radiation energy
because recall that though the Newtonian supermassive stars have
$p_r \gg p_g$, they still have $\rho_r \ll \rho_g$. On the other
hand, since Eq.(31) demands $\rho_r \gg \rho_g $ at appropriately
high $z$, such an occurrence
 actually automatically denotes domination of radiation pressure.
Therefore, for $x \gg 1$, we may rewrite Eq.(38) as
 \b
 x \approx ~ z ~\l({M \over 15 M_\odot}\r)^{1/2} \approx0.25 ~z\l({M
\over  M_\odot}\r)^{1/2} \qquad z\gg 1 \e The absence of $z$ in
Eq.(24) demands that one can have $x \gg 1$ only for very large
values of $M$. On the other hand, the presence of $z$ in Eq.(39)
shows that, in the extreme relativistic case, one can have RPSSs
{\em for an arbitrary stellar mass}, high or low. Hence as one would
proceed towards $z \to \infty$ during continued gravitational
collapse, one must obtain radiation pressure supported
configurations ($x \gg 1)$ at arbitrary mass rather than finite mass
BHs. Although we worked only in the regions of $z \ll 1$ (Eq.[24])
and $z \gg 1$ (Eq.[39]), the close similarity in the forms of
Eqs.(24) and (39) strongly suggests that Eq.(39) may be valid for
RPSSs in the entire relativistic range range of $z >1$.
\section{Discussion}
Newtonian stars are defined by $z \ll 1$ as well as $\rho_r \ll \rho_g$, i.e., $y \ll 1$. It is such self-imposed constraints which cause $x \ll 1$.
 However, even in the Newtonian regime, i.e., despite having $y \ll 1$
 and $z \ll 1$, one can have RPSSs ($x \gg 1$) for
masses above $7200 M_\odot$.  Hence we find that the non-occurrence
of low mass RPSSs in the Newtonian regime is intricately linked with
the occurrence of $z \ll 1$ and $y \ll 1$ in such cases. On the
other hand, GR is unleashed in full glory during continued
gravitational collapse when one can have $z \gg 1$ and $y \gg 1$ for
an {\em arbitrary mass}, low or high. It naturally followed then
that, in the extreme GR case, one can have a RPSS even at arbitrary
low $M$. While occurrence of Newtonian supermassive stars may be
only a theoretical possibility, Einstein RPSSs must be a reality
because of the following simple reason:

As $z$ tends to increase indefinitely during continued collapse, the
strong gravity almost completely traps the collapse generated
neutrinos and photons within the body of the star\citep{b9}. The
density of trapped radiation also increases because of stellar
matter -radiation interaction, i.e., the diffusion of the internal
radiation\citep{b4}. As a result, the trapped radiation pressure and
energy density increases at least as fast as $p_r, ~\rho_r \sim
R_0^{-3} (1+z)^{2}$\citep{b9}. On the other hand, the locally
defined Eddington luminosity grows as\citep{b10}
 \b
  L_{ed} = {4 \pi G M
c\over \kappa} (1+z) \e where $\kappa \approx 0.4$ cm$^2$ g$^{-1}$
is the Thomson opacity. The radiation pressure associated with
$L_{ed}$ is
 \b p_{ed} = {1\over 3} {L_{ed} \over 4 \pi R^2  c}
\propto R^{-2} (1+z)
 \e
  Therefore, $p_r /p_{ed} \propto R^{-1} (1+z)$.
Thus as $R$ decreases and $1+z$ increases $p_r \to p_{ed}$
eventhough, initially, $p_r \ll p_{ed}$ and $L \ll L_{ed}$. It is at
this stage that $L \to L_{ed}$ and $\alpha \to 1$. Simultaneously,
one will have both $p_r \gg p_g$ and $\rho_r \gg \rho_g$ at this
stage. Specific models of GR continued collapse which {\em do not}
restrict $\rho_r$ by any means {\em do clearly show that repulsive
effects} of unabatedly rising of $\rho_r$ and $p_r$ may not only
counterbalance the inward pull of gravity but even cause the fluid
to bounce back\citep{b11, b12}. Such studies are fully consistent
with the generic picture of formation quasistatic GR RPSSs at
arbitrary mass scale discussed in this paper. The observed
luminosity of such RPSSs will however be lower by a factor of
$(1+z)^2$  than the local value because of joint effect of
gravitational redshift and gravitational time dilation:
 \b
 L_{ed}^\infty = {L_{ed} \over
(1+z)^2} = {4 \pi G M c\over \kappa (1+z)}
 \e
Consequently, even if the system would be assumed to be at a given
fixed value of $z=z$, its observed duration as seen by a far away
astronomer would be
 \b
  t(z) = { Mc^2 \over
L_{ed}^\infty} = {k c \over 4 \pi G} (1+z)
 \e
  Since in principle,
during continued collapse, $z \to \infty$, clearly, the observed
time scale for depletion of mass energy becomes infinite for
arbitrary value of the opacity $\kappa$: $
 t =\infty $. Hence the RPSSs tend to collapse for infinite duration
 in order to attain the BH ($z=\infty$) state and, therefore, may be
 called as ``Eternally Collapsing Objects'' (ECOs)\citep{b13, b14}.
  It has been also shown that
 since the eventual BH mass would
  be zero, the comoving proper time for its formation would also be
 infinite\citep{b14}. Since the observed BH Candidates must
 be formed in gravitational collapse and of finite age, they must be
 ECOs ($z \gg 1$) rather than true BHs ($z=\infty$).
In retrospect, long back the
 RPSSs were suggested as the central engine of quasars\citep{b1,
 b2}. However this attempt failed because such RPSSs (i) were
 considered to be either Newtonian or Post Newtonian objects with low temperatures,
  (ii) The basic
 source of energy liberation was considered to be of nuclear origin.
 In contrast the relativistic RPSSs considered here are fed by
 energy release due to secular gravitational contraction and the
 source of energy is the entire mass energy ($E= M c^2$). Even if
 they would momentarily be unstable, the contraction generated
 luminosity would ensure that they pass from one quasistatic state of
 $z=z_1 \gg 1$ to another with $z_2  > z_1$.
\section{Acknowledgements}
The author thanks Stanley Robertson, Darryl Leiter,  Norman
Glendenning and P.S. Negi for various  discussions.  The author is
also grateful to Felix Aharonian for persistent encouragement. The
anonymous referee is also thanked for some constructive suggestions.
 %\end{document}


\begin{thebibliography}{99}
\bibitem[\protect\citeauthoryear{Chandrasekhar}{1967}]{b7} Chandrasekhar, S., 1967 {\it An Introduction to the Study of Stellar Structure}
(Dover, New York, 1967)
\bibitem[\protect\citeauthoryear{Fowler}{1966}]{b2} Fowler, W.A.,
1966, Astrophys. J., 144, 180
\bibitem[\protect\citeauthoryear{Herrera and Santos}{2004}]{b11} Herrera, L. \& Santos, N.O.,
2004, Phys. Rev. D70, 084004
\bibitem[\protect\citeauthoryear{Herrera, Prisco \& Barreto}{2006}]{b12} Herrera, L., Prisco, A.D., and
Barreto, W., 2006, Phys. Rev. D73, 024008 (gr-qc/0512032)
\bibitem[\protect\citeauthoryear{Hoyle and Fowler}{1963}]{b1} Hoyle, F. \&
Fowler, W.A., 1963, 125, 169
\bibitem[\protect\citeauthoryear{Mitra}{1998}]{b10} Mitra, A., 1998,
preprint (astro-ph/9811402)
\bibitem[\protect\citeauthoryear{Mitra}{2000}]{b13} Mitra, A., 2000,
Found. Phys. Lett., 13(6), 543, (astro-ph/9910408)
\bibitem[\protect\citeauthoryear{Mitra}{2002}]{b14} Mitra, A., 2002,
 Found. Phys. Lett., 15, 439, (astro-ph/0207056)
\bibitem[\protect\citeauthoryear{Mitra1}{2005}]{b5} Mitra, A., 2005a,
preprint, (gr-qc/0512006)
\bibitem[\protect\citeauthoryear{Mitra2}{2005}]{b6} Mitra, A., 2005b,
preprint, (physics/0504076)

\bibitem[\protect\citeauthoryear{Mitra}{2006}]{b4} Mitra, A., 2006, Mon. Not. Roy. Astron.
Soc. Lett., 367, L66 (gr-qc/0601025)

\bibitem[\protect\citeauthoryear{Mitra and  Glendenning}{2006}]{b9} Mitra. A \&
Gendenning, N.K., 2006, Preprint
%\bibitem[\protect\citeauthoryear{Tooper}{1964}]{b8} Tooper, R.F.,  Astrophys. J., 1964, 140, 434
%\bibitem[\protect\citeauthoryear{Bowers \& Deeming}{1984}]{b3}
%Bowers, R.L. and Deeming, T., 1984, {\it Astrophysics I, Stars} (Jones and %Bartlett, Boston, 1984)
%\bibitem[\protect\citeauthoryear{Mitra}{1998}]{b2} Mitra, A., 1998, astro-ph/9811402

%\bibitem[\protect\citeauthoryear{Shapiro \& Teu}{1983}] {b3}  Shapiro, S. and  Teukolsky, S.A., 1983, {\it Black Holes, White Dwarfs, and
%Neutron Stars: The Physics of Compact Objects}, (Wiley, New York, 1983)



\bibitem[\protect\citeauthoryear{Vaidya}{1951}]{b8}  Vaidya, P.C., 1951, Proc. Ind. Acad. Sc., A33, 264






%\bibitem[\protect\citeauthoryear{LR}{2003}]{b15} Leiter, D. \&
%Robertson, S.L., 2003, Found. Phys. Lett., 16, 143,
%(astro-ph/0111421)

\bibitem[\protect\citeauthoryear{Weinberg}{1972}]{b3}   Weinberg, W., 1972, {\it Gravitation and Cosmology: Principles and
Applications of General Theory of Relativity},  (John Wiley, New
York, 1972)
\end{thebibliography}
\end{document}